\documentclass{article}
\usepackage{graphicx}  
\usepackage{graphics}
\usepackage{subfigure}
\usepackage{amsmath}
\begin{document}
%-----------------------------------------------------------------------------------------------------------------------------------
\date{}
\title{The Spatial Variability of Vehicle Densities as Determinant of Urban Network Capacity}
\author{Amin Mazloumian$^1$, Nikolas Geroliminis$^2$, Dirk Helbing$^1$ \\ 
 $^1$ETH Zurich, UNO D11, Universit\"atstr. 41, 8092 Z\"{u}rich, Switzerland \\
$^2$Urban Transport Systems Laboratory, EPF Lausanne, \\ GC C2 389, Station 18, 1015 Lausanne, Switzerland}
%-----------------------------------------------------------------------------------------------------------------------------------
\date{}
\label{firstpage}
\maketitle
\date{}
%-----------------------------------------------------------------------------------------------------------------------------------
\begin{abstract}{Urban traffic, network dynamics, congestion spreading, macroscopic traffic model, fundamental diagram, route assignment}
Due to the complexity of the traffic flow dynamics in urban road networks, most quantitative descriptions of city traffic so far are based on computer simulations. This contribution pursues a macroscopic
(fluid-dynamic) simulation approach, which facilitates a simple simulation of congestion spreading in cities. First, we show that a quantization of the macroscopic turning flows into units of single vehicles is necessary to obtain realistic fluctuations in the traffic variables, and how this can be implemented in a fluid-dynamic model.
Then, we propose a new method to simulate destination flows without the requirement of individual route assignments. Combining both methods allows us to study a variety of different simulation scenarios. These reveal fundamental relationships between the average flow, the average density, and the {\it variability} of the vehicle densities. Considering the inhomogeneity of traffic as an independent variable can eliminate the scattering of congested flow measurements.  
The variability also turns out to be a key variable of urban traffic performance.
Our results can be explained through the number of full links of the road network, and approximated by a simple analytical formula.
\end{abstract}
%-----------------------------------------------------------------------------------------------------------------------------------
%-----------------------------------------------------------------------------------------------------------------------------------
%-----------------------------------------------------------------------------------------------------------------------------------
%-----------------------------------------------------------------------------------------------------------------------------------
\section{Introduction}
%-----------------------------------------------------------------------------------------------------------------------------------
%-----------------------------------------------------------------------------------------------------------------------------------
Traffic flow theory has relied for decades on fundamental laws, some of which were inspired by analogies with fluid flows, electrical currents and the like. These laws are usually based on fundamental indicators (speed, density and flow) and describe how they are linked together. The most common relationship is called the "fundamental diagram", according to which the flow increases with the vehicle density up to the capacity of the road, and then it decreases down to zero, when the road is  congested. This diagram is used to characterize the "regimes" of traffic flow (free or congested) in \emph{a specific road location or one link}, and it was first studied by Greenshields (1935). Designed primarily for  highway traffic flows, such laws are not sufficient to describe the entire complexity of traffic flows in urban road networks. Besides, the flows show a significant scattering, especially in the congested regime (see for example Kerner 1996; Tu 2008). Nevertheless, by aggregating the highly scattered plots of flow vs. density from individual fixed detectors, it was empirically found for the city of Yokohama that a Macroscopic Fundamental Diagram (MFD) with low scattering exists, which links space-mean flow and density (Geroliminis \& Daganzo 2008).

The first instance of a macroscopic fundamental diagram (MFD) showing an optimum car density was presented by Godfrey (1969). Earlier studies  looked for macro-scale traffic patterns in data of lightly congested real-world networks (Godfrey 1969; Ardekani \& Herman 1987; Olszewski \textit{et al.} 1991) or in data from simulations with artificial routing rules and static demand (Mahmassani \textit{et al.} 1987; William \textit{et al.} 1987; Mahmassani \& Peeta 1993). However, the data from all these studies were too sparse or not investigated deeply enough to demonstrate the existence of an invariant MFD for real urban networks. Support for its existence has been  given only very recently (Geroliminis \& Daganzo 2007, 2008). These references  showed that 
\begin{itemize}
\item the MFD is a property of the network itself (infrastructure and control) and not of the demand, i.e. the MFD should have a well-defined maximum and remain invariant when the demand changes both with the time-of-day and across days and  
\item the space-mean flow is maximum for the same value of critical vehicle density, independently of the origin-destination tables. 
\end{itemize}
To evaluate topological or control-related changes of the network flows (e.g.  due to a re-timing of the traffic signals or a change in infrastructure), 
Daganzo \& Geroliminis (2008) and Helbing (2009) have derived analytical theories for the urban fundamental diagram, using a density-based and a utilization-based approach respectively. Curves derived from both theories fit the data obtained from the Yokohama experiment well (Daganzo \& Geroliminis 2008;  Helbing 2009).

Despite these and other recent findings supporting the existence of well-defined macroscopic fundamental diagrams (MFDs) for urban areas, it is not obvious whether the MFDs would be universal or network-specific. More real-world experiments are needed to identify the types of networks and demand conditions, for which invariant MFD's with low scatter are found. Daganzo (2007) argued that if the traffic conditions change slowly with time, a MFD should exist for networks with a homogeneous spatial distribution of congestion. However, Buisson \&  Ladier (2009) showed with real data from a medium-size French city that heterogeneity has a strong impact on the shape/scatter of a MFD, that may not even remind of a MFD in some cases, e.g. for freeway networks under non-recurrent conditions.

Congestion in urban traffic networks is by nature unevenly distributed in space. This is because of spatial inhomogeneity  
\begin{itemize}
\item in demand (some parts of the network attract or generate more trips than others), 
\item in road infrastructure (some routes in the network have more lane-miles) and 
\item in control (different types, among them traffic signals and stop signs, and different control settings within each type, e.g. offsets or green times of successive signals).
\end{itemize}
Both, in empirical data and computer simulations, it has been observed that traffic conditions may significantly vary for similar travel activities and traffic volumes. That is, one day's traffic dynamics can be characterized by widespread and long-lasting congestion, while traffic flow is barely or not at all affected on other days, despite similar
origin-destination flows (Bernard \textit{et al.} 2007).
It is likely that this property does not only follow from the network structure, but is also a consequence of instabilities of traffic flows (for an overview
see, e.g., Helbing 2001; Helbing \& Moussaid 2009; Helbing \textit{et al.} 2009). 

To address such questions, it is common to use computer simulations of urban traffic flows (Axhausen \& G\"{a}rling 1992; Biham \textit{et al.} 1992; Nagel \& Schreckenberg 1992; 
Schadschneider \& Schrenkenberg 1993; Daganzo 1994; Herrmann 1996; Klar \& Wegener 1998; Charypar \& Nagel 2005; Herty \textit{et al.} 2006; Bretti \textit{et al.} 2007; L\"{a}mmer \textit{et al.} 2007
; Ma \& Lebacque 2007; De Martino 2009; Padberg \textit{et al.} 2009). However, the frequency and evolution of flow breakdowns and congestion spreading processes are still poorly understood, probably because of the interplay between topology and dynamics (Zhao \textit{et al.} 2005). This applies both to the local level where congestion originates, as well as to the network level in terms of how it spreads.

In this paper, we are investigating how the inhomogeneity in the spatial distribution of car density affects the shape, scatter and even the existence of a macroscopic relation between the average flow and vehicle density in urban networks. As data availability from cities is limited, we are following a simulation-based approach to study a range of scenarios. 
The main contributions of this paper are (i) the introduction of innovative modeling techniques, namely (i) a macroscopic flow quantization, (ii) a memoryless traffic flow routing, and (iii) a better understanding of the urban-scale macroscopic fundamental diagram. Our routing method does not require origin-destination tables and complicated routing decisions or route assignment, which are necessary in most urban microsimulation models. In addition, by applying the flow quantization, we are able to reproduce a realistic variability of network flows even for the same average car density. Moreover, we discover the variability as a key variable of urban traffic flows, which reveals clear functional relationships rather than producing large data clouds for congested traffic.

Our results emphasize that the spatial aggregation of traffic variables cannot guarantee a well-defined relationship between the average density and flow, especially when the network is congested. We observe that for the same average density of vehicles in the network and the same assumptions regarding the origin-destination flows, there is a wide variation of possible average network flows, potentially even ranging from free flow to gridlock. A key component in all cases is the spatial variability of congestion at a 
specific time, as expressed by the standard deviation of density among all links. The degree of spatial inhomogeneity is highly correlated with the number of full links in the network. 
%Full links create spillbacks, which block the output of upstream adjacent links and significantly decrease the average speed and flow of the system. 
 Each full link prevents upstream links from discharging vehicles. 

After describing the main modeling components of our simulator in \S2, we will start in \S3 with the investigation of a macroscopic fundamental diagram for networks with invariant density over time without trip generation or trip termination. Vehicles are moving randomly in the network and turn in a memoryless way. We will show that urban network capacity is not a deterministic quantity and investigate the reasons of the variability. In \S4 we will extend the analysis in networks, where a percentage of trips is directed towards a center of attraction and trips are generated at a rate that keeps the average vehicle density constant. In \S5 we will analyze more general networks with time-dependent densities and multiple destinations. Finally, \S6 will provide a summary, discussion and outlook.

%-----------------------------------------------------------------------------------------------------------------------------------
%-----------------------------------------------------------------------------------------------------------------------------------
%-----------------------------------------------------------------------------------------------------------------------------------
\section{Model}
%-----------------------------------------------------------------------------------------------------------------------------------
%%
%%
%%
\begin{figure}[ht]
\centering
\includegraphics[width =50mm]{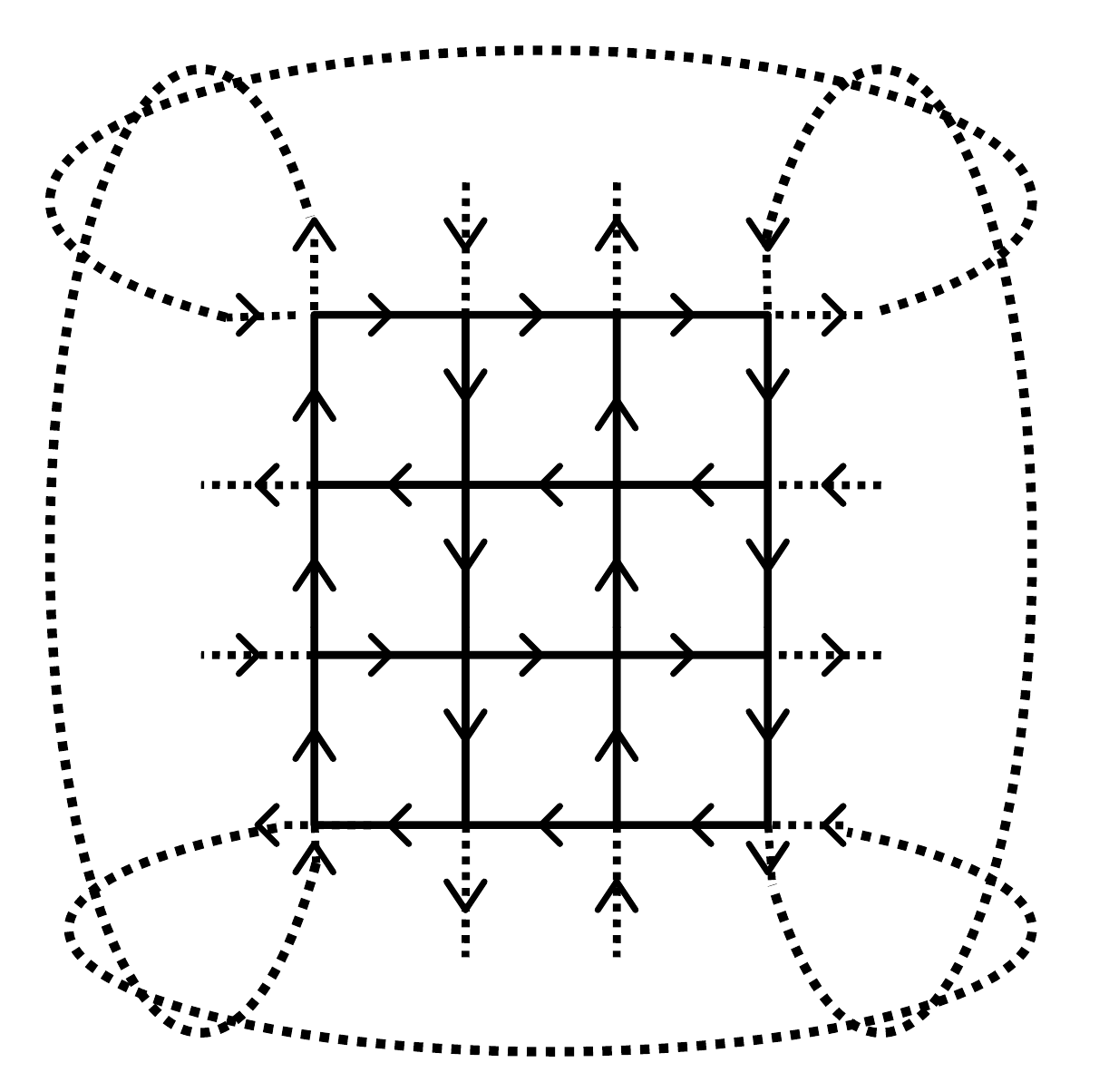}
\caption{In our computer simulations, a city-center is represented by a lattice-like uni-directional road network reminding of traffic flow in Manhattan or Barcelona. 
Intersections are controlled by fixed-cycle traffic signals. Arrows show the flow directions. Vehicles leaving the network at the boundary enter the network again 
at the opposite side, i.e. we assume periodic boundary conditions.  
}\label{fig:schematic}
\end{figure}
We model the road network of a city-center by a $30\times30$ lattice network (see figure \ref{fig:schematic}). 
The links of the network represent one-way road sections, and the nodes represent intersections.
To obtain a better control over the number of vehicles in the network, we assume periodic boundary conditions, i.e. we connect the intersections at the boundary to the 
intersections located on the opposite side. 
Moreover, all road sections have the same length $L$ of $200$ metres and a two-phase fixed-cycle traffic signal setting with the same green time period $g$ of $30$ seconds and an amber plus red time period of $36$ seconds (i.e. a cycle length $C$ of $66$ seconds). 
As drivers have different driving characteristics, their response to perfectly time-coordinated signals is not ideal. Therefore, we added some stochasticity to traffic light offsets, to imitate adaptive green wave schedules. 
 The offset at each signal is randomly selected from a uniform distribution with an average, which is equal to the value that creates a ``green wave'' and minimizes the delays during light traffic conditions. That is, on average the green phase at a downstream intersection starts $L/V^0$ seconds after the one upstream, where $V^0$ is the free flow speed.

\subsection{Underlying Dynamics}\label{dynamics}
The dynamics of the traffic flow in the road network is determined by the section-based traffic model (Helbing, 2003; Helbing \textit{et al.} 2007).
The model exhibits significant features of traffic flows such as the conservation of vehicles, jam formation, and spillovers.

\begin{figure}[ht]
\centering
\includegraphics[width =100mm]{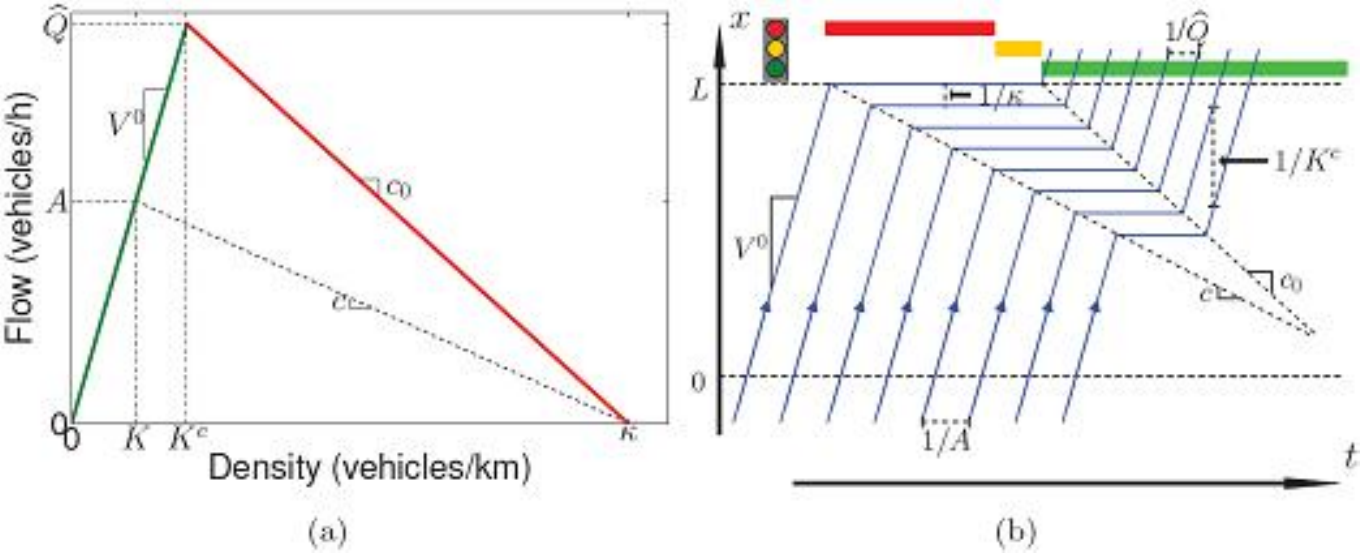}
\caption{(a) The fundamental diagram defines a flow-density relation for individual road sections. In the triangular approximation of the fundamental diagram, vehicles move with the free speed $V^0$ below the critical density $K^c$. When the density $K$ is greater than $K^c$, a vehicle queue forms, and the speed $c$ 
of the upstream moving congestion front is given by the arrival
flow. The dissolution speed $c_0  < 0$ of congested traffic, i.e. the speed of downstream congestion front, 
 is a characteristic constant. (b) Schematic illustration of vehicle trajectories on a road section of length $L$. Vehicles move forward at the free speed $V^0$ or are stopped in a vehicle queue
(horizontal lines), which forms during the amber and red time
periods behind the traffic light (located at $x=L$). 
}\label{fig:schematics}
\end{figure}

 As illustrated in figure \ref{fig:schematics}, the fundamental diagram along individual road sections is approximated by a triangle with two characteristic speeds:
 the desired speed of vehicles or speed limit $V^0$, and the resolution speed $c_0$ of traffic jams (Helbing 2003). For any road section, the model calculates the temporal evolution of the arrival flow $A(t)$, the departure flow $O(t)$, 
and the location $l(t)$ of the upstream jam front, considering the ``permeability'' 
$\gamma(t)$ (reflecting the traffic signal) and the turning factors $\alpha_i(t)$. More specifically, the turning factor $\alpha_i(t)$ indicates what fraction of the outflow
is turning left or right, depending on the respective intersection (see figure \ref{fig:schematic}). Due to vehicle conservation, $1-\alpha_i(t)$ is the fraction of vehicles 
moving straight ahead into the next downstream road section.
%----------------------------------------------------------------------------------------
%----------------------------------------------------------------------------------------
%----------------------------------------------------------------------------------------
%----------------------------------------------------------------------------------------
%----------------------------------------------------------------------------------------
%%
%%
%----------------------------------------------------------------------------------------
%----------------------------------------------------------------------------------------
%----------------------------------------------------------------------------------------
%----------------------------------------------------------------------------------------
%----------------------------------------------------------------------------------------
%----------------------------------------------------------------------------------------
The maximum arrival flow of a road section is limited by the maximum flow $\widehat{Q}$, if the road section is not fully congested. 
it is given by its departure flow at time $t-L/c$, where $L$ denotes the length of the road section:
\begin{equation}
0\leq A(t) \leq \widehat{A}(t)= 
\begin{cases} 
\widehat{Q} & \text{if $l(t)<L$,}
\\
O(t-L/c) &\text{if $l(t)=L$.}
\end{cases}
\end{equation}
Similarly, the departure flow of a road section is bounded to the maximum flow $\widehat{Q}$, when there are some delayed 
vehicles $\Delta N > 0$. But when no vehicle is delayed by congestion, the departure flow is given by the arrival flow at time $t-L/V^0$, where $L/V^0$ is the free 
travel time along the road section:
\begin{equation}
0\leq O(t) \leq \widehat{O}(t)= \gamma(t)
\begin{cases} 
A(t-L/V^0)& \text{if $\Delta N(t)=0$,}
\\
\widehat{Q} &\text{if $\Delta N(t)\neq0$.}
\end{cases}
\end{equation}
The number of delayed vehicles $\Delta N(t)$ in the above equation evolves according to
\begin{equation}
\frac{d\Delta N(t)}{d t}=A(t-L/V^0)-O(t).
\end{equation}
In order to represent a traffic light at an intersection serving two flows, the permeability $\gamma(t)$ of the served road 
section is set to $1$ during the green phase, while the permeability of the other one is set to $0$ when the traffic light is red. Apparently, during switching intervals (amber times), both  permeabilities are set to $0$.    

After computing the maximum arrival and departure flows $\widehat{A}(t)$ and  $\widehat{O}(t)$, one can determine the actual departure flow of a road section (which determines the actual flows of the two following road sections). This is done by
 restricting it to the maximum arrival flows of the two following road sections. 

In our computer simulation, we consider a triangular fundamental diagram with characteristic speeds $V^0=50\text{3km/h}$ and $c_0=-14.28\text{ km/h}$, and a maximum
density $\kappa$ of $140$ vehicles/km.

\subsection{Flow Quantization}

\begin{figure}[ht]
\centering
\includegraphics[width =100mm]{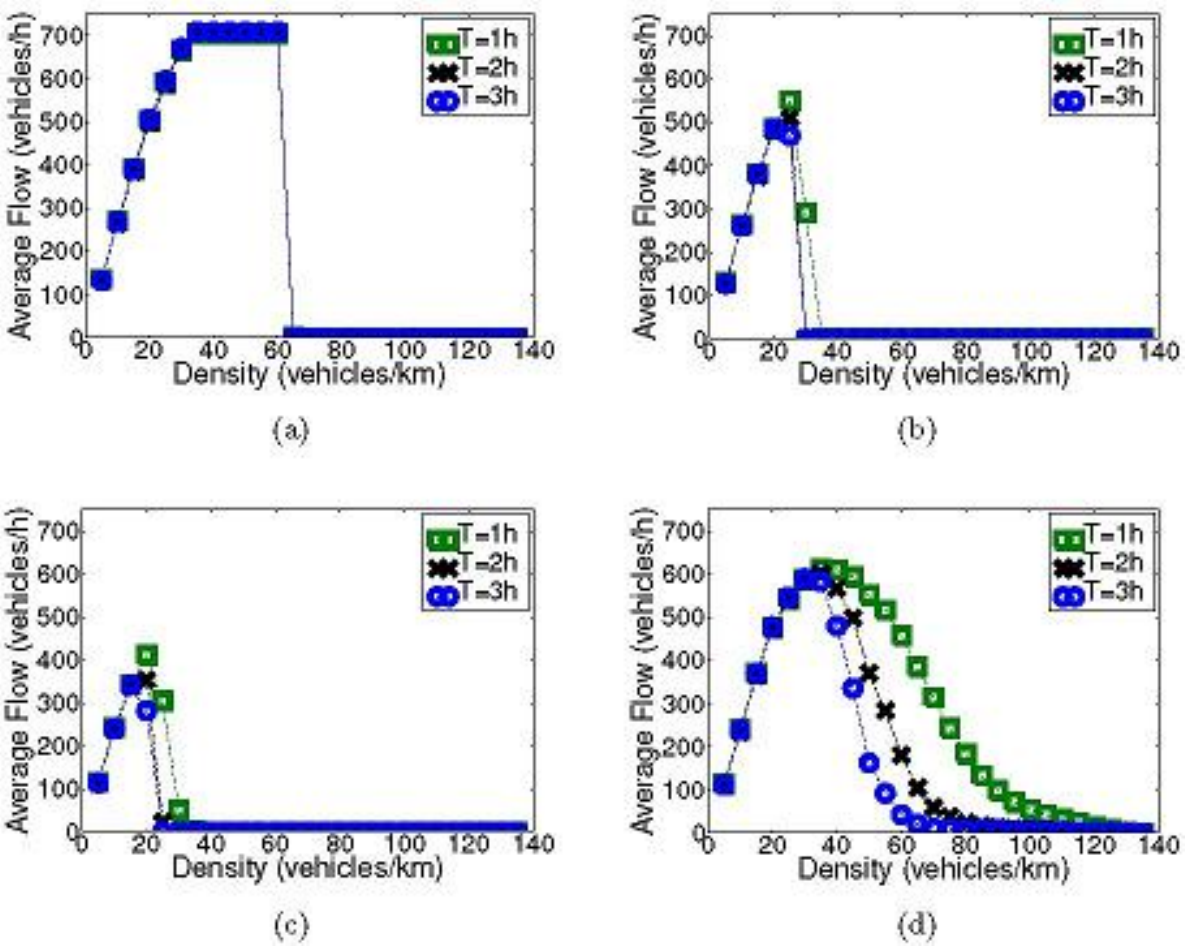}
\caption{ (a)-(c) Changing the turning factors of outflows with a fixed frequency leads to a 
sharp transition between the free flow and the gridlock state. However, this large sensitivity to the number of vehicles in 
the network is unrealistic. In (a), outflows are evenly and deterministically distributed among
road sections. In (b), the turning factors change with a low variability ($\alpha_i(t)\in[0.25,0.75]$). 
In (c), the turning factors change with a high variability ($\alpha_i(t)\in[0,1]$). (d) By quantization of the outflows, i.e. random determination of the 
turning decisions of each single vehicle, a realistic macroscopic fundamental diagram is obtained. If the vehicle density is higher than $30$ vehicles/km, the average network flow drops as time passes.
}
\label{fig:fdx}
\end{figure}
In the previous section, we explained the dependency of the arrival and departure flows of road sections on the 
time-dependent turning factors $\alpha_i$ at intersections. Here, we assess the temporal dynamics of 
the turning factors. The first question is: On what time scale do we have to model the change of turning factors? At first glance, the cycle time of traffic lights
seems to be a good choice. To test this, we have simulated a uniform demand in our network by changing the turning factors at
the beginning of each service cycle. For example, one may 
assume equal probabilities of flows to continue straight or turn at all intersections. Then, the averages of the turning factors 
$\alpha_1(t)$ and $\alpha_2(t)$ at any intersection would be $0.5$ over a long enough interval of time. 
We have studied three variants of turning factors:  

\begin{enumerate}
\item{Deterministic: Outflows of all road sections are time-independent and evenly distributed 
among the following road sections $\alpha_i=0.5$.}
\item{Random with low variability: For any road section, in the beginning of each traffic cycle, the turning factor $\alpha_i(t)$
is chosen uniformly at random in the interval $[0.25,0.75]$.}
\item{Random with high variability: The turning factor $\alpha_i(t)$ of any road section is chosen uniformly at random in the interval $[0,1]$ 
in the beginning of each service cycle.}
\end{enumerate}

We have simulated $500$ realizations for each of the three cases and for a range of fixed average densities.
The results of the simulation are illustrated in figure \ref{fig:fdx}, which shows the macroscopic fundamental 
diagrams, i.e. the relationship between the average outflow of road sections (the network flow)
and the number of vehicles in the network (the network density). The simulations start with uniform distances among the vehicles and equal accumulation in every road section. We measure simulation time after a transient period of half an hour for a time period of three hours. A large number of simulation runs are performed for different initial densities. For all three specifications of time-dependent turning factors, there is a sharp transition between the free flow and the gridlock state. This,  however, does not seem to be realistic. 

To deal with turning factors, we propose a new method based on flow discretization, as conjectured by Helbing (2005):
\begin{enumerate}
\item{In agreement with microscopic models, it is assumed that the outflow does not simultaneously take both directions. We rather assume that the outflow either turns or continues straight ahead, corresponding to binary values $\{0,1\}$ of $\alpha_i(t)$.}
\item{The flow direction is decided for each equivalent of one vehicle, i.e. we quantisize the outflow into units of single vehicles.} 
\item{The outflow of any road section may be interrupted because of a red light or the spillover of the road section in the chosen direction. 
In both cases, the decision for the direction of outflow does not change, and the considered road section is blocked, i.e. vehicles cannot leave the road section anymore for some time.} 
\end{enumerate}
To be comparable with the previous analysis of constant average turning frequencies, vehicles choose their flow direction by tossing a fair coin. As shown in figure \ref{fig:fdx}(d), our method leads to 
a realistic macroscopic fundamental diagram for uniform travel demand: We observe that, even if the turning factors are balanced, if the initial distribution of traffic is homogeneous, and if the traffic signals are controlled in the same way, congestion does not distribute homogeneously in the network, as is well-known from reality. 
In \S \ref{section:mfd}, we analyze the properties of the resulting macroscopic
fundamental diagram in more detail. 

\subsection{Route-Choice}

Besides developing a new flow quantization technique, we have so far described, how a uniform demand in a network can be simulated.
However, a dominant factor for the formation of urban-scale congestion is the inhomogeneous distribution of traffic demand
throughout the network. For instance, during commuting hours, there is a higher density of traffic around workplaces.
Therefore, we need a mechanism to direct pre-defined fractions of flows to destination areas in order to simulate inhomogeneous demand. This goal is not trivial, as
our model does not include origin-destination tables for vehicles. We will therefore propose now a simple routing protocol capable of 
simulating scenarios with multiple destination areas and inhomogeneous traffic demands. Despite its simplicity, our method approximates a shortest-path type of route-choice 
towards destination areas with a minimum number of turns at intersections. 

For illustration let us consider two types of vehicles. The first type represents vehicles with homogeneous demand throughout the network, while the second type steer towards specific destination areas. 
However, since we do not distinguish individual vehicles from each other, route-choices cannot be assigned to vehicles. Assuming that vehicles of both types are   
distributed everywhere in the network, our problem reads as follows: How can a fraction of outflow of any road section be routed to a specific destination area? 
By periodic assignment of routes to the outflow of road sections proportionally to the demand of the destination areas, the problem translates into the following one: At each intersection, in which direction does a vehicle with a certain destination area drive? To be specific, if $20\%$ of vehicles in the network are routed to a specific destination area, $1$ out
of $5$ vehicles of the outflow of any road section is routed towards the destination area, while the other $4$ are routed to other destination areas or choose direction randomly in our simulation.

\begin{figure}[ht]
\centering
\includegraphics[width=100mm]{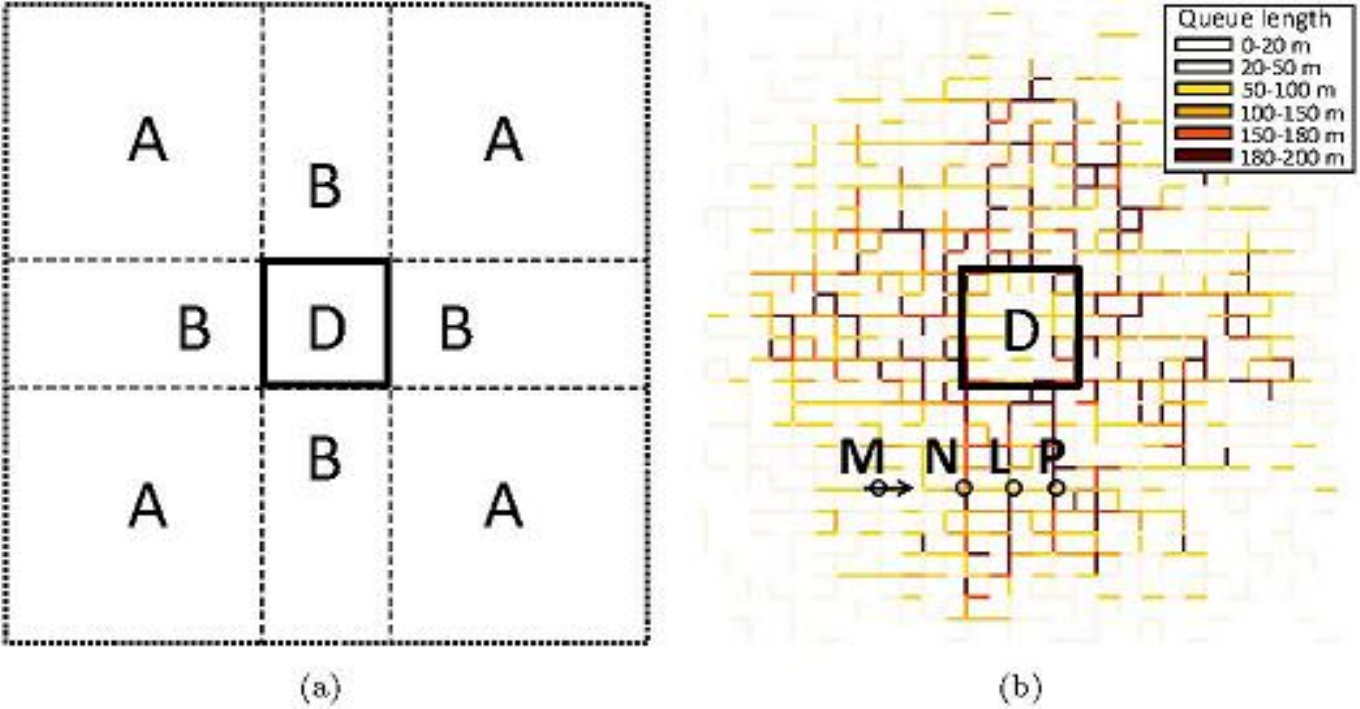}
\caption{(a) A destination area divides our network into $9$ regions. We propose a shortest-path type of route-choice 
towards destination areas with a minimum number of turns at intersections. The underlying rule is to drive straight, unless the Manhattan distance to the specified 
destination area increases. With this route-choice, as shown in (b), a  vehicle that drives horizontally towards intersection $M$ continues straight to
intersection $N$. Then, the vehicle has three choices to turn towards the destination area, namely at intersections $N$, $L$ and $P$.
At intersection $N$, the destination is assumed to be in the two left columns of the destination area with 
probability $\frac{1}{3}$. Hence, the vehicle chooses the straight direction with probability $\frac{2}{3}$ and turns towards the destination area with probability
$\frac{1}{3}$. The next intersection at which the vehicle can turn towards the destination area is $L$. Reasoning in a similar way, 
the vehicle chooses to go straight with probability $\frac{1}{3}$ and to turn towards $D$ with probability $\frac{2}{3}$.
If did not turn towards $D$ at intersections $N$ and $L$, it certainly turns towards the destination area at intersection $P$, as driving straight would prolong its route.
}   
\label{fig:routing}
\end{figure}

Consider a destination area $D$ with the shape of a square and let the network be partitioned
into $9$ regions, as marked by different letters in figure \ref{fig:routing}(a). The simulation scenario consists of $4$ regions in the corners (marked by $A$), $4$ regions adjacent to the 
boundaries of the destination area (marked by $B$), and the destination area itself (marked by $D$). To reach the destination area $D$, 
we can apply the following rule of thumb: \emph{Drive straight, unless the Manhattan distance to the specified destination area increases.} Assuming
an ordered numbering of the rows and columns of the network, the Manhattan distance between two intersections is obtained by summing up the absolute values of their row difference and
column difference. The minimum distance that a vehicle can traverse to reach the destination area is at least its 
minimum Manhattan distance to the boundaries of the destination area. This is due to the fact that vehicles move parallel or orthogonal to the boundaries. Note that the trip length can be
larger by the length of one road section for vehicles moving in the opposite direction of the destination area, which requires a turn at the nearest intersection.   
 
There are slight variations of our routing rule in different regions:
\begin{itemize}
\item{Vehicles that drive in the corner regions (marked by $A$ in figure \ref{fig:routing}(a)) only turn if driving straight increases their Manhattan distance, while turning decreases their distance to the destination area. By distance we mean the minimum Manhattan distance to the boundaries of the destination area. All vehicles in this region drive straight until they reach a neighbouring region (marked by $B$ in figure \ref{fig:routing}(b)).}
\item{Vehicles driving in the neighbouring regions (marked by $B$ in figure \ref{fig:routing}(a)), drive straight, if their direction
of movement is orthogonal to the boundaries. Otherwise, if a vehicle is driving parallel to the boundaries, driving straight decreases its distance 
to some of the intersections in the destination area and increases its distance to those it passes in parallel. In this case, we set the probability 
of turning towards the destination area proportional to the number of nodes in the destination area for which the distance is decreased by turning (see figure \ref{fig:routing} for a concrete example). In this way, for any vehicle that drives from an $A$-region to a $B$-region toward a $n \times n$ destination area, there are $n/2$ choices to turn towards the destination area. Therefore, 
the probability to turn towards the destination area is $2/n$ at its first intersection in $B$,  while the probability of turning is $1$ if it reaches the other boundary. 
}
\item{Vehicles that reach the destination area choose their direction randomly to simulate a homogeneous demand.}
\item{At the intersections on the boundary of the network, vehicles choose their direction randomly. A vehicle that leaves the network from one side, enters it again from the other side.}
\end{itemize}

In figure \ref{fig:routing}(b), $20\%$ of vehicles are routed towards the destination area. Vehicles with destination $D$ turn towards their destination just in $B$-regions, in which the direction towards the destination area is chosen with a higher probability. Therefore, $B$-regions are more congested than $A$-regions. 
%-----------------------------------------------------------------------------------------------------------------------------------
\section{Macroscopic Fundamental Diagram}\label{section:mfd}
Let us denote by $q_{i}$ and $n_{i}$ the flow and the number of vehicles on link $i$ during the time period of a signal cycle. We are interested in the aggregate patterns produced by these variables at the network level. To this end, we define the average network flow by 
\begin{equation}
Q=\sum_i\frac{q_{i}}{ML}, 
\end{equation}
and the average network density of vehicles by
\begin{equation}
K= \sum_i \frac{n_{i}}{ML},
\end{equation}
where $M$ is the total number of links in the network and $L$ is the length of links ($0.2$ km in our simulations). To analyze the effect of spatial heterogeneity, we represent the global standard deviation of the number of vehicles among all links in the network by $S$ and the number of full links in the network by $F$. A full link is related with high levels of congestion, because it blocks the departures from upstream links and thereby significantly decreases the upstream vehicle flow. 
Each full link blocks the outflow of the two upstream links. Thus, we approximate the number of blocked links by twice the number of full links, $2F$. 

\begin{figure}[ht]
\centering
\includegraphics[width =90mm]{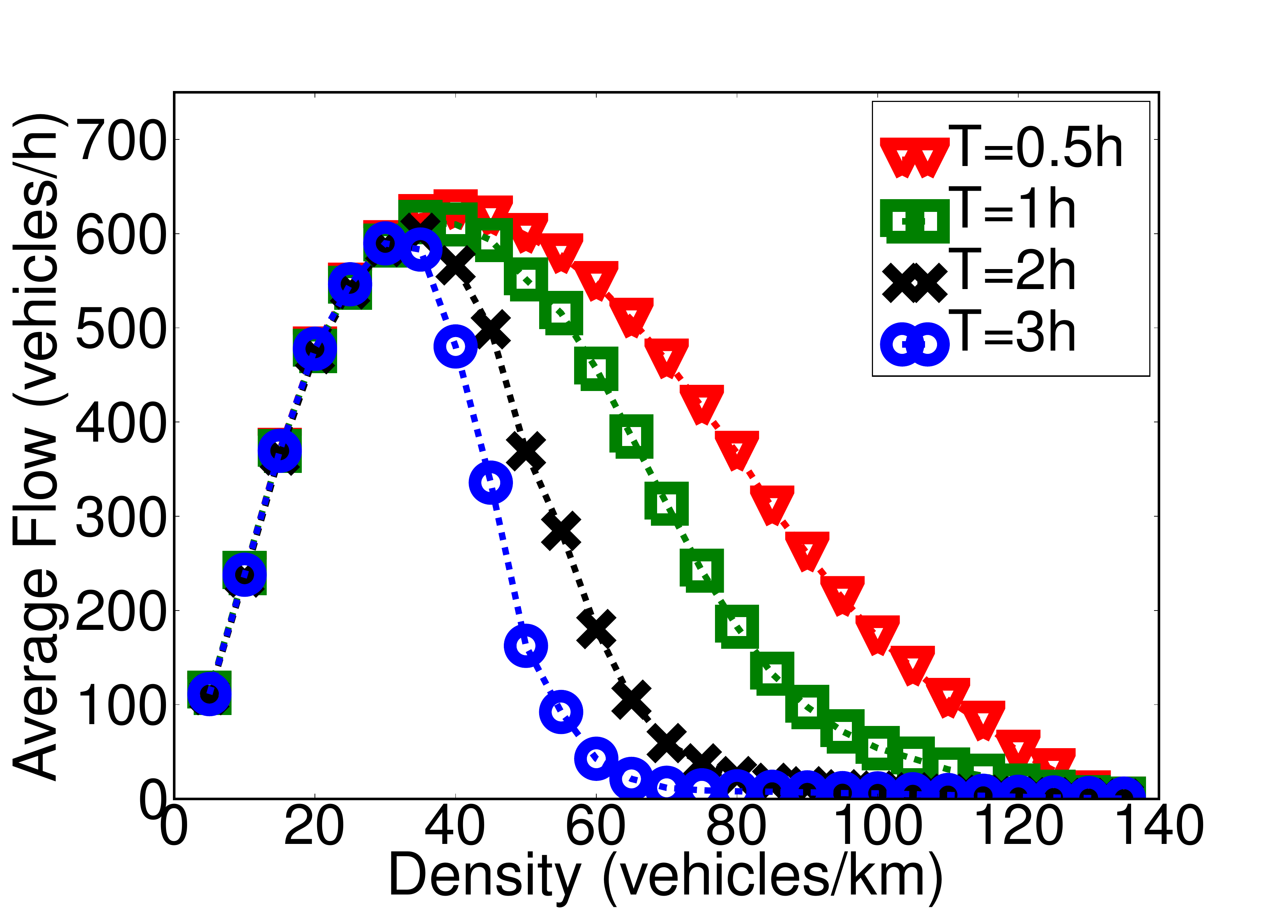}
\caption{
For a wide range of network densities $K$, we run $500$ simulations for each density over $3$ hours and calculate the average network flow $Q$ in all runs at half an hour, one hour, two and three hours of simulation time. Note that there is no invariant Macroscopic Fundamental Diagram (MFD) for the network, as the flow $Q$ decreases significantly with time for $K\ge40$ vehicles/km. The curves for $T=1$, $2$, $3$ h are the same as in figure \ref{fig:fdx}(d).
}\label{fig:fdour}
\end{figure}

Let us now study the aggregate relation between the network flows $Q$ and the densities $K$ in the different simulation runs. First, we simulate a network with an invariant number of vehicles moving randomly over the network and turning at intersections with different probabilities. We are interested to identify whether the performance of the network as expressed by the average flow $Q$ is a predetermined or a varying quantity over time when the number of vehicles moving in the network is constant. In other words, we investigate whether there exists a well-defined Macroscopic Fundamental Diagram $Q=Q(K)$.  To obtain these measurements, for a wide range of average densities with values between $20$ vehicles/km and $120$ vehicles/km, we run a large number of simulations ($500$ runs for each density) and calculate the average network flow of all runs for the same average density after half an hour, one hour, two and three hours of simulation. The results are summarized in figure \ref{fig:fdour}.
 It turns out that the variability of flows for low or very high densities is negligible. For low values of the average network density (less than $30$ vehicles/km), the network flows stabilize at a characteristic value which is invariant over time and varies little among simulation runs. Note that the stabilization threshold of $30$ vehicles/km is slightly smaller than the value of the  network density that maximizes the network flow ($35$ vehicles/km). For very high values of density (greater than $120$ vehicles/km), the network reaches a state of gridlock very quickly, and congestion cannot be dissolved. However, for intermediate values of the network density, we observe a high level of variability. For example, for an average density of $60$ vehicles/km, the flow decreases from $500$ vehicles/h after one hour of simulation to $250$ vehicles/h after $2$ hours and less than $100$ vehicles/h after $3$ hours. It is clearly visible,   that the network consistently leads to smaller flow as time passes. Note also that, for densities higher than $70$ vehicles/h, the network reaches a state of gridlock after $2$ hours in the great majority of runs.

This high variability of network flows, especially when the network density is at the critical value that (sometimes) maximizes flow, deserves further investigation in the following and implies important questions: Why does the traffic situation significantly vary from one day to another even if travel demand  is similar? How often should we expect a traffic collapse and congestion spreading? What variables would facilitate a better description of the severity of traffic congestion? Can we obtain any functional relationship between the capacity and other key variables rather than having a large scattering describing congested traffic phenomena?

%%-----------------------------
\begin{figure}[ht]
\centering
\includegraphics[width=126mm]{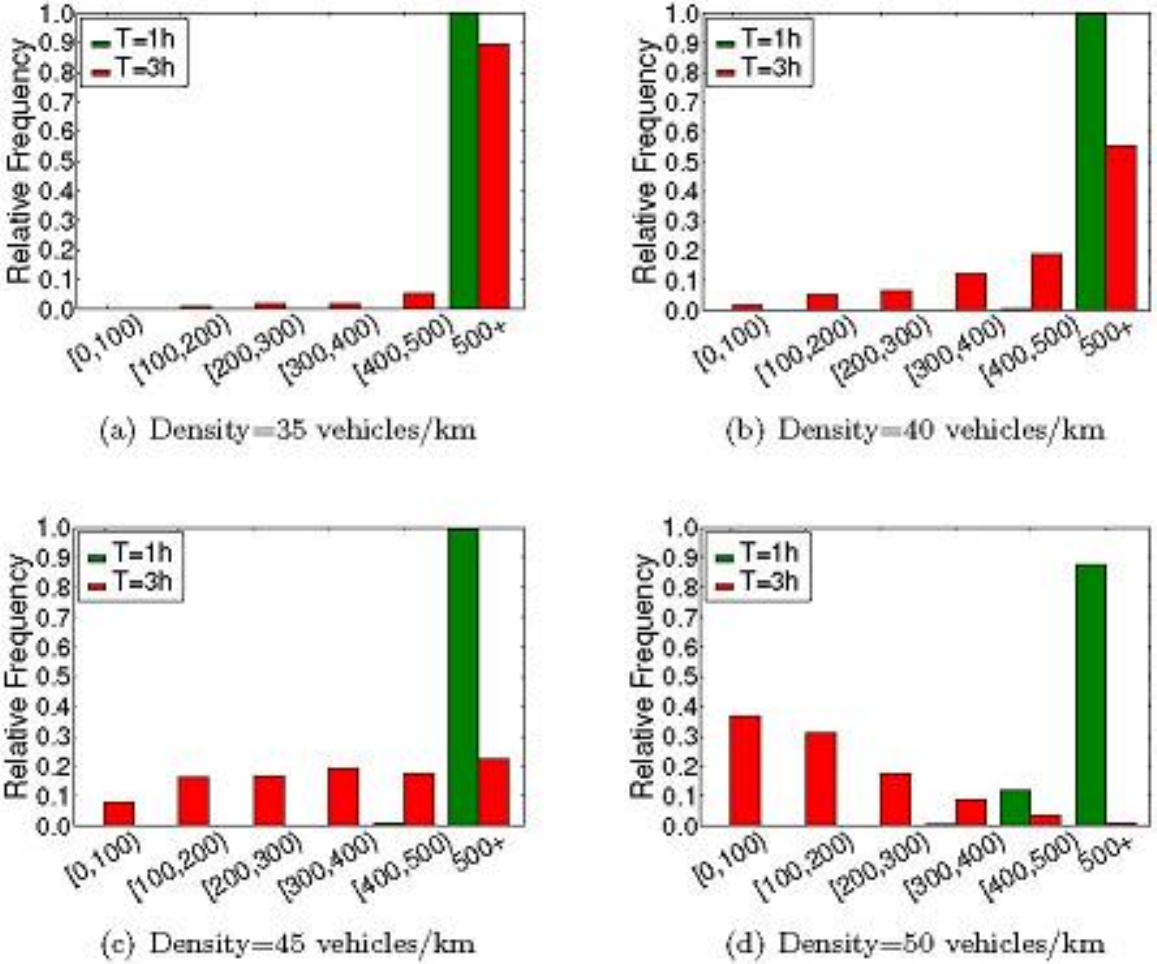}
\caption{Histograms of the average network flows $Q$ for network densities of $35$, $40$, $45$ and $50$ vehicles/km after a time period $T=1$ and $3$ hours of simulation time. Each graph is the result of $500$ simulation runs. Variability of $Q$ is negligible for $K=35$ vehicles/km, but significant for higher $K$. Flow $Q$  varies between $0$ and maximum flow for a wide range of $K$ and $T$. 
}
\label{fig:freqxxxx}
\end{figure}
%%-----------------------------
To learn more in this direction, we further analyze the aforementioned simulation data and plot histograms of the network flows at different times during the computer simulation for a range of different densities. These histograms shed more light on the density range for which the variability of flow is crucial. Figure \ref{fig:freqxxxx} shows flow histograms for network densities of $35$, $40$, $45$ and $50$ vehicles/km after $1$ and $3$ hours of simulation time. The density values are commonly observed in congested city centers worldwide. While the probability of network failure (as reflected by flows much smaller than capacity) is negligible for a network density of $35$ vehicles/h, this probability is significant for higher densities. For example, we observe an almost uniform distribution of flows between zero and $600$ vehicles/h for a density of $45$ vehicles/km after $3$ hours of simulation, while for a network density of $50$ vehicles/km, the histogram is skewed left. Given the fact that congested periods in city centers last long (e.g. the speed in the city center of Yokohama remains less than $8$ km/h for about $5$ hours each weekday, (Geroliminis \& Daganzo 2008), an explanation of the empirically observed flow variability would provide useful insights to develop more efficient control strategies in the future. These results also suggest that macroscopic fundamental diagrams should be network specific. 
\begin{figure}[ht]
\centering
\includegraphics[width =90mm]{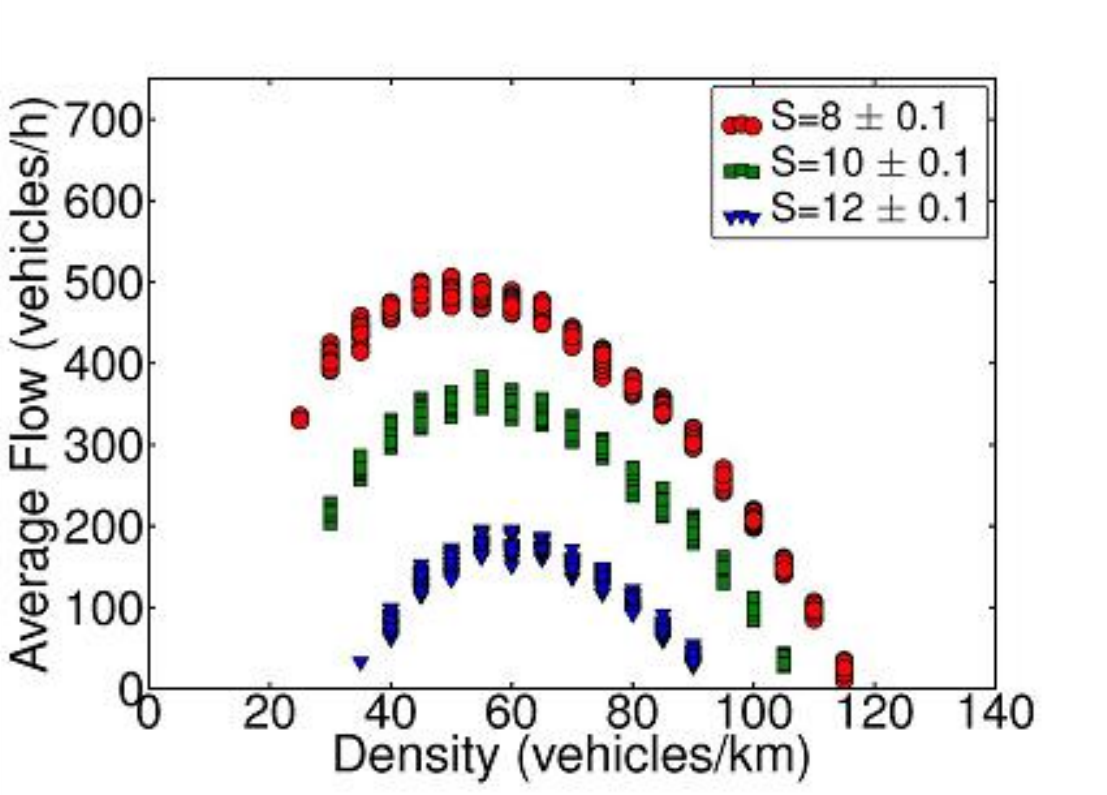}
\caption{
Relationship between the average network flow and the average network density, when the data are distinguished according to the standard deviation $S$ of vehicles in the different road
sections. The large reduction of the scattering of congested traffic data as compared to figure \ref{fig:fdour} suggests to consider $S$ as a relevant variable of urban traffic flow.
}
\label{fig:fdourv}
\end{figure}

We now show that, when considering the spatial distribution of congestion, which is reflected by the standard deviation $S$ of the number of vehicles in all links in the network, we can obtain functional relationships, even in cases where the  MFDs show a large amount of scattering. To this end, in each simulation run we select only the signal cycles for which $S$ is within a range of $\pm0.1$ vehicles/link around some selected values, e.g. $8$, $10$ and $12$ vehicles/link. We then plot $Q$ vs. $K$ for all times during which the standard deviation is within the predefined range. The results are shown in figure \ref{fig:fdourv}. We observe that the variability in $Q$ can be explained through the standard deviation $S$. The remaining scattering of the flow values is less than $50$ vehicles/h in all cases. While these results are very encouraging, they cannot be directly utilized to develop control strategies because (i) $S$ is a time-dependent quantity and (ii) the critical density that maximizes flow varies with $S$.

\begin{figure}[ht]
\centering
\includegraphics[width=90mm]{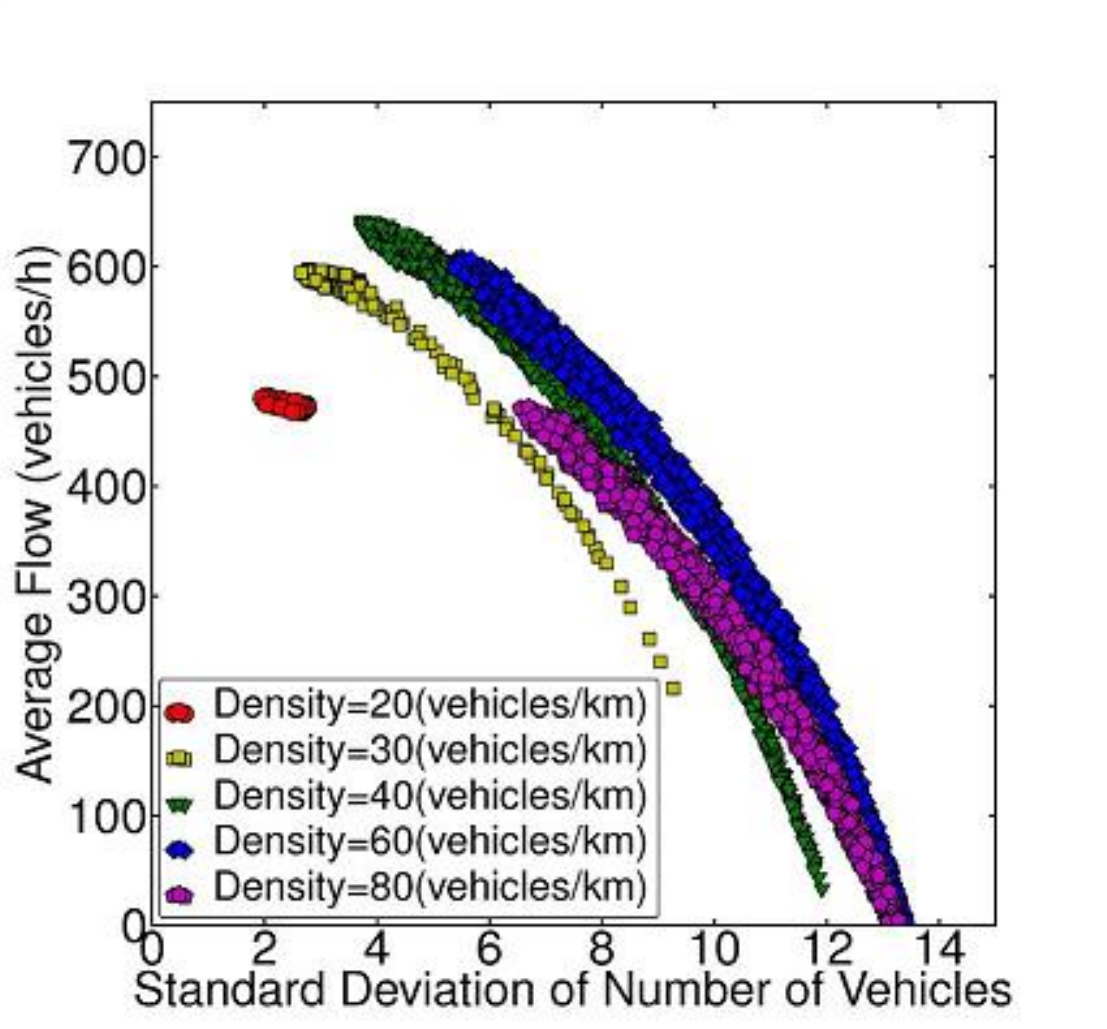}
\caption{Average vehicle flow as a function of the variability $S$ of the number of vehicles in the different road sections for various values of the average density. For the average vehicle  density $K = 20$ vehicles/km,
the standard deviation $S$  of number of vehicles in the different road sections and the network flow is high. For the average density $K = 30$ vehicles/km, which is slightly smaller than the critical density, the average flow $Q$ is close to
$600$ vehicles/h for small $S$. The flow drops below this value only in $5\%$ of the simulation runs. However, for higher values of $K$, the average flow varies significantly and reaches gridlock 
for high values of $S$.}
\label{fig:fvuniform}
\end{figure}

It is intuitively clear that a high standard deviation in the spatial distribution of vehicles  is connected with lower values of network flows for the same average network density. But, is there a functional relationship between the average flow $Q$ and the variability $S$?  The answer is "yes", as can be seen in figure \ref{fig:fvuniform}. We observe that, for a given value $K$ of the average density, there is a unique monotonously falling relationship between the average flow $Q$ and the variability $S$ in all simulation runs, and the remaining scattering of the flow data is low. 
  For small densities the flow never breaks down, while it may reach gridlock for higher densities, as the variability $S$ increases.
%For $K=20$ vehicles/km, $S$ is always small and the flow never breaks down. For $K=30$ vehicles/km, which is a value slightly smaller than the critical density, $Q$ is close to $600$ vehicles/h for small values of $S$, but it drops bellow this value with a small probability, when the vehicle density in the network becomes inhomogeneous (i.e. when the variability $S$ increases). For higher values of $K$, the flow varies significantly and reaches gridlock, as $S$ assumes high values. 

It is not difficult to estimate the maximum value of $S$ for a given vehicle density $K$. It occurs when almost all vehicles are in fully occupied links that create spillbacks and do not allow any outflow. During gridlock (with almost zero flow), the number of full links can be approximated as     $F^{\max} \approx \frac{MK}{\kappa}$, where $\kappa$ denotes the maximum density and $M$ is the total number of links in the network. The standard deviation of the number of vehicles per link can be approximated as 
\begin{equation}
S^{\max} \approx \sqrt{F^{\max}[(K-\kappa)L]^2+(M-F^{\max})(KL)^2}. 
\end{equation}
For $K=40$, $60$ and $80$ vehicles/km, we find $S^{\max}=12.7$, $13.9$, $13.9$ vehicles/link respectively, which is very close to the values obtained in our simulations. 
It is easy to show that $S^{\max}$ becomes maximum for $K=\frac{\kappa}{2}=70$ vehicles/km. 

So far, we have shown that, for a system with periodic boundary conditions and homogeneous turning rates, the number of full links is a key variable of the system.  It is therefore interesting to ask whether there is any functional relationship between $Q$ and $F$ in more general cases?
%%
%%
%-----------------------------
\begin{figure}[ht]
\centering
\includegraphics[width =90mm]{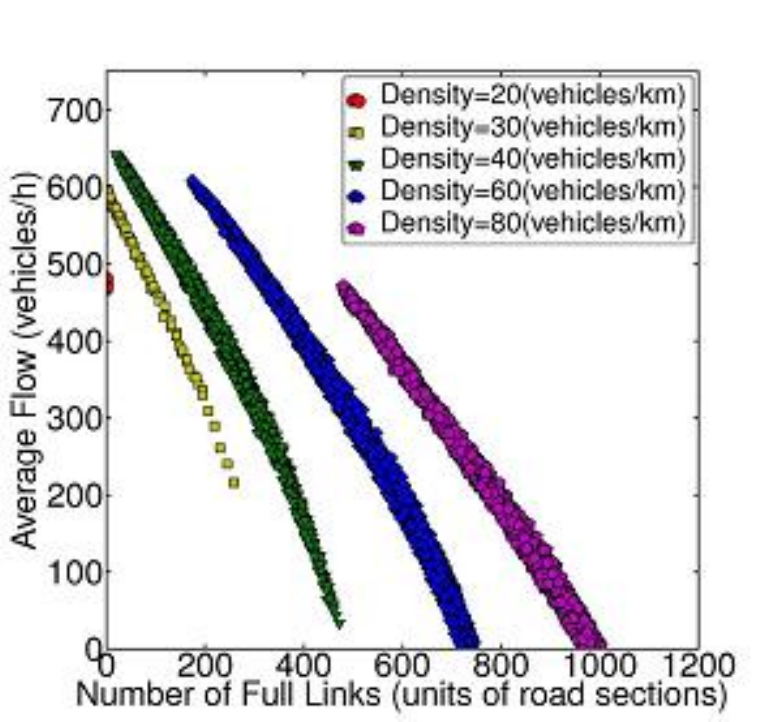}
\caption{The relationship between the network flow and the number of full links in the network is almost linear, because uncongested parts of network are not much affected by the 
congested parts. At lower densities, vehicles do not necessarily stop at every signal and can travel more than one link per cycle, which results in a larger decrease of the flow by each full link in the network.}
\label{fig:fsuniform}
\end{figure}
%-----------------------------
%%
%%
To answer this question, we determine for each simulation run the number of full links within the period of one signal cycle. As the flow and the number of vehicles in each link are  continuous variables, we assume that a link is full when the number of vehicles is greater than $98\%$ of the maximum possible value, $\kappa L$. Figure \ref{fig:fsuniform} shows $(Q, F)$ pairs from all simulation runs for different densities $K$. The results show consistent and almost linear relationships with a low degree of scattering and similar slopes, which slightly decrease as $K$ increases. The values of $F^{\max}$ for different densities $K$, as estimated before, match the simulated data well. 

The numerical findings can be understood as follows: For high values of vehicle density $K$, most links without constraints by downstream queues operate at capacity. In contrast,  full links   create spillbacks and decrease the flow to zero for the two upstream links connected with them (remember that the simulated network has only one-way road sections). Thus, the average flow decreases as 
\begin{equation}
\frac{d Q}{d F}\approx 2\left(\frac{g}{C}\right)\frac{\widehat{Q}}{M}=0.8 \text{ vehicles/h} 
\end{equation}
per additional full link, where $g$ is the duration of the green phase during each cycle $C$, $\widehat{Q}$ is the maximum flow, and $M$ is the total number of links in the network. This value is approximately equal with the value of the slope for $K=80$ vehicles/km. For smaller densities, the slope is expected to be higher, because delays at traffic signals are smaller, and vehicles do not necessarily stop at every signal, i.e. they can travel more than one link per cycle.

%-----------------------------------------------------------------------------------------------------------------------------------
%-----------------------------------------------------------------------------------------------------------------------------------
%-----------------------------------------------------------------------------------------------------------------------------------
%-----------------------------------------------------------------------------------------------------------------------------------
\section{Simulations with Trip Generation and Termination}

\begin{figure}[ht]
\centering
\includegraphics[width=126mm]{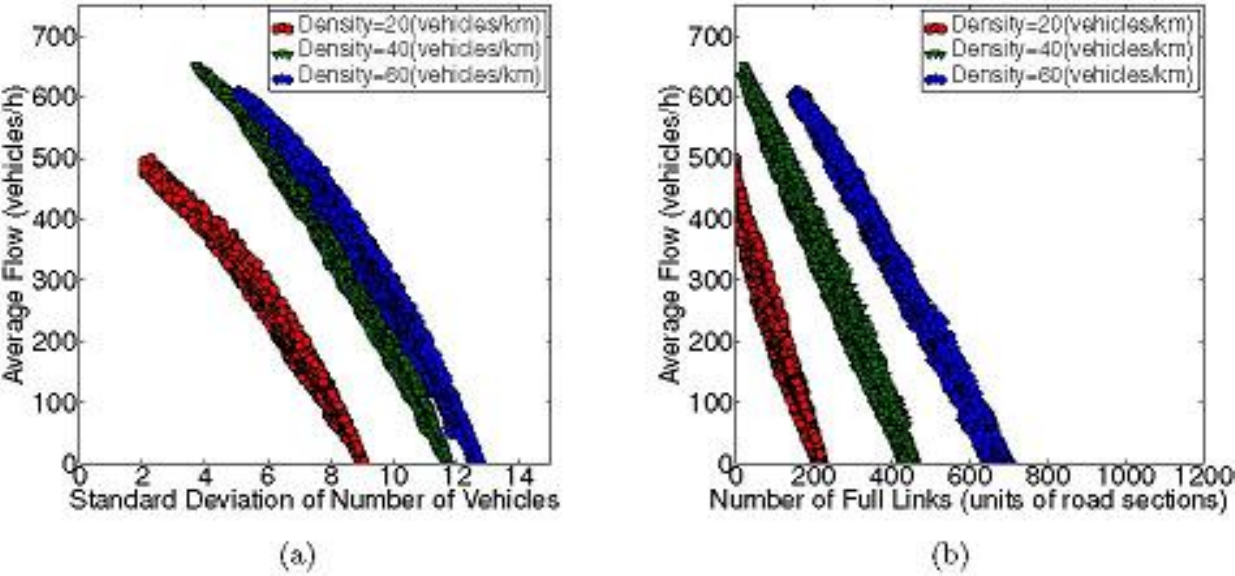}
\caption{
Traffic simulation with a $1\text{3km}\times 1\text{3km}$ destination area located in the center of a $6\text{ km}\times 6\text{ km}$  large network with periodic boundary conditions.  
The data points correspond to the scenarios where fractions of $5$, $10$, $15$, $20$, $25$, and $50$ percent 
of the vehicles are routed to the destination area. Trips are terminated inside the destination area. There is a uniform trip generation throughout the network that keeps
the vehicle density constant. Besides the vehicle density in the network (which was kept constant) both, (a) the standard deviation of number of
vehicles among links and (b) the number of full links, are determinants for the average network 
flow, even when the demand is not uniformly distributed in the network. 
}
\label{fig:onecenter}
\end{figure}

So far, our analysis of the macroscopic fundamental diagram (MFD) has revealed that, for a periodic road network with constant vehicle density, the average flow is determined 
by the average density and the standard deviation of vehicles (variability) or the number of full links with a high accuracy (see figure \ref{fig:fsuniform}, \ref{fig:fvuniform}). The results of the previous section are independent of origin-destination tables. If these results remain valid in more general traffic scenarios, they could provide a convincing explanation of  the  
mysterious variability of congestion from one day to another despite identical travel demands. 
We have observed that, even when the computer
simulation starts with a homogeneous distribution of vehicles, the same traffic signal setting, and a uniform demand, many different spatial distribution of congestion occur due to the randomized turning of vehicles.   
However, would this also be true if the spatial distribution of congestion was more predictable? We will now simulate such traffic conditions by directing a fraction of vehicles in the network towards a destination area, so that the vehicles concentrate around the destination area. Will the standard deviation of the number of vehicles among links still a determinant of  the average network flow when the demand is non-homogeneous? We find that gridlock occurs within a short time also when a small fraction of vehicles are routed to a destination area. This is because of the fast accumulation of vehicles in the road sections
around and inside the destination area. To remove vehicles from the links in the destination area and free them for vehicles that want to enter it, we still have to model trip termination. 
In this section, we will keep the number of vehicles in the network constant. This is implemented  by adding a vehicle to a random link (trip generation) for each removed vehicle in the destination area. In the next section, we will finally study scenarios with a varying number of vehicles in the network. 

We model trip termination by setting a fixed probability of removal of vehicles for the road sections inside the destination area. The removal is made from the quantisized outflow from the road sections. This implies that no vehicle is removed from a blocked road section, as in reality. If vehicles with different destinations are well-mixed with the same ratio  everywhere, there is the following relationship between the fraction of vehicles which are routed to the destination area and the trip termination probability: \emph{If the fraction $r$ of the vehicles in the network are routed to the destination area, the fraction $1-r$ of the total inflow of the destination area should leave the destination area.} This, gives us the possibility to calculate the trip termination probability $p$ as a function of the fraction $r$ and the size of the destination area. The probability that a vehicle enters and leaves the destination area after traversing $n$ road sections is $(1-p)^n$. Therefore, the average of the probability of leaving the destination area is 
\begin{equation}\label{walker}
\overline{(1-p)^n}=1-r,
\end{equation} 
where the average is performed over many vehicles with different path lengths $n$. As the path length $n$
 is independent of the travel times along the road sections, and the termination probability $p$ can be numerically calculated by simulating random walkers which traverse one link per step. 
To check whether equation (\ref{walker}) is satisfied, we measure the path lengths of many random walkers. More specifically, each random walker starts a trip from a random intersection on the boundary of the destination area, chooses random directions at intersections, and traverses links until it leaves the boundary again. Our results show that, for a $5\times 5$ destination area, if $1$ vehicle out of $u$ vehicles from the outflow of the road sections is routed to the destination area ($r=\frac{1}{u}$), $1$ vehicle out of $5u-5$ terminates its trip inside the destination area ($p=\frac{1}{5u-5}$). For instance, when $20\%$ of vehicles are routed to a destination area of size $1\text{3km}\times 1\text{3km}$, the trip termination probablity at intersections inside the destination area is $0.05$. Figure \ref{fig:onecenter} 
illustrates the results. For each vehicle density, we simulated $20$ realizations with $u \in \{2,3,4,5,10,20\}$. The good match between figure \ref{fig:fvuniform}, \ref{fig:fsuniform}, and   \ref{fig:onecenter} verifies that, both the standard deviation of number of vehicles among links and the number of full links determine the average network flow together with the vehicle density in the network, even if the travel demand is not uniformly distributed in the network. 
%-----------------------------------------------------------------------------------------------------------------------------------
\section{Time-Dependent Scenarios}

\begin{figure}[ht]
\centering
\includegraphics[width=126mm]{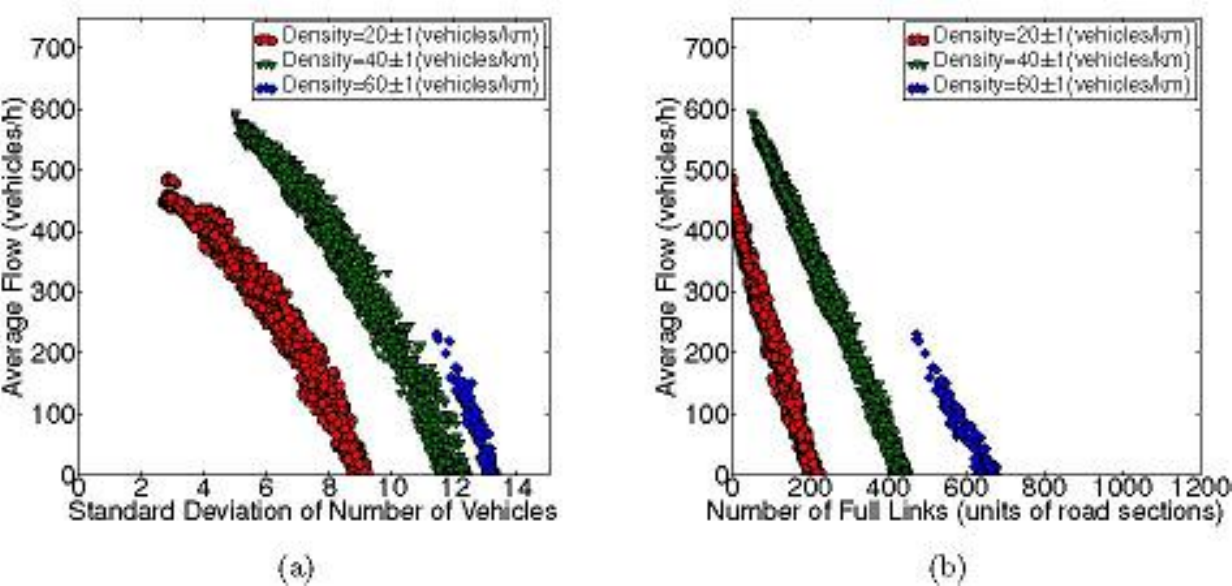}
\caption{
Even for varying traffic volumes, the average network flow can be expressed as a function of the average vehicle density and (a) the standard deviation of number of vehicles , or (b) 
the number of full links.
}
\label{fig:varyload}
\end{figure}

In reality, urban road networks are subject to varying network densities. However, all simulations in the previous sections kept the number of vehicles in the network  constant by a balanced trip generation and termination. Hence, as final test we check whether our findings stay the same for time-varying traffic volumes. 
To represent an open system, we simulate a fast growing demand in the network.
For this, we simulate 
 scenarios in which trip generation is independent of trip termination. Our simulation varies from fast trip generation ($5$ vehicles/link/h) to very fast trip generation ($10$ vehicles/link/h), and from small fractions of vehicles being routed to the destination area (e.g. $5\%$) to large fractions (e.g. $50\%$). To compare our results with those presented in figures \ref{fig:fvuniform}, \ref{fig:fsuniform}, and \ref{fig:onecenter}, from all the different simulations, we 
choose the data points for which the average vehicle density is $K=20\pm1$, $40\pm1$, and $60\pm1$ vehicles/km. 
The good agreement with previous results, as illustrated in figure \ref{fig:varyload}, supports that even under varying traffic volumes, the average network flow can be expressed as a function of the standard deviation of the number of vehicles and the   average vehicle density, independent of time-dependent patterns of travel demand. Due to fast growing demand, we observe that for the average  network density of $K=60$ vehicles/km network flows are always low.

%-----------------------------------------------------------------------------------------------------------------------------------
\section{Summary, Conclusions, and Outlook}
We have studied how the spatial variability of vehicle density can affect the traffic performance at the network level. Most studies until now have looked at macroscopic relationships between the average flow and density, but not at the relevance of their variability. The results of this paper show that the  standard deviation of density is a key variable which is  required (i) for the existence of an invariant urban macroscopic fundamental diagram, (ii) to explain the wide variation of average network flows (potentially ranging from high values up to gridlock even in case of the same average density and demand), (iii) to provide a robust and well-defined macroscopic functional relationship even in cases where origin-destination flows significantly vary.  A simple explanation of these dependencies is that an inhomogeneity in the spatial distribution of car density increases the probability of a spillover, which substantially decreases the network flow. 

To reach the above conclusions, we have introduced new modeling techniques, namely a flow quantization and a memoryless traffic flow routing. 
 Flow quantization is natural in micro-simulation (car-following or cellular automata models), but non-standard in macro-simulation models. It was not known before that this consideration of a micro-feature would make such a big difference in macro-models. This allowed us to reproduce a realistic variability of network flows in fluid-dynamic computer simulations even for the same average car density, and without the need of detailed origin-destination tables and complicated routing assignment models.  

While the results of this paper provide a clearer understanding of traffic in cities, further investigations are needed to identify  (i) whether these functional relationships also hold for more complex road networks and turning relations, (ii) how traffic congestion spreads with time as a function of topological and demand characteristics, 
and (iii) how these outcomes can be applied to real cities in order to avoid high levels of congestion. For freeways, one needs to investigate the existence of an invariant macroscopic fundamental diagram, as the flow-density relationship and traffic dynamics are different (e.g. Kerner 1998;  Orosz \textit{et al.} 2009).

To enhance traffic performance, we are interested in strategies that can reduce the variability of the vehicle densities. For this, one has to decrease the number of fully occupied links in a network, e.g. by prioritizing critical vehicle queues (Helbing \& L\"{a}mmer 2007; L\"{a}mmer \& Helbing 2008) or restricting access to neighbourhoods which exceed certain density thresholds (Daganzo 2007). Recently, Helbing \&  Mazloumian (2009)  proposed a  signal control, which explains the slower-is-faster effect when the utilization of a road section is so small that it requires extra time to collect enough vehicles for an efficient service during the green phase. Similarly, the slower-is-faster strategy would suggest to restrict the inflow to congested areas to keep the service capacity high (Geroliminis \& Daganzo 2007; Johansson \& Helbing 2008).

Another future direction is the simulation of more complex traffic networks. This could include irregular networks with varying number of lanes, link lengths and counterflows, allowing to turn left and right at each intersection, and finally multi-centric cities with multiple attraction centers. Interesting phenomena/questions arise (i) when pockets for left turn movements spill over and block through movements on the same link or (ii) previously separated congestion areas reach each other, thereby causing a serious large-scale collapse of traffic flow. Also, additional studies are needed to further understand the impact of network topology and its interplay with the network dynamics and congestion spreading. Finally, as simulations invariably involve untested assumptions, these should be tested by extensive real-life experiments. Such empirical studies will reveal, whether the functional dependence of the average network flow on the average density and the number of full links are independent of different demand profiles, as suggested, which would be very useful for theoretical and practical considerations.
\\
\emph{
AM and DH are grateful for support by the Gottlieb Daimler and Karl Benz Foundation within the framework of their BioLogistics project and by the ETH Competence Center ``Coping with Crises in Complex Socio-Economic Systems'' (CCSS) through the ETH Research Grant CH1-0108-2 as well as funds from the ETH Foundation. \\
NG is grateful for the great hospitality, the excellent working conditions and the valuable collaboration at the Chair of Sociology, in particular of Modeling and Simulation, ETH Zurich.
}
%-----------------------------------------------------------------------------------------------------------------------------------
\begin{enumerate}
\item{Ardekani, S. \& Herman, R. 1987 Urban network-wide traffic variables and their relations. \textit{Transportation Science} \textbf{21} (1), 1--16.}
\item{Axhausen, K.W. \& G\"{a}rling, T. 1992 Activity-based approaches to travel analysis: conceptual frameworks, models, and research problems. \textit{Transport Reviews}, \textbf{12} (4), 323--341.}
\item{Bernard, M. \& Axhausen, K.W. 2007 A highway design concept based on probabilistic operational reliability. \textit{11th World Conference on Transportation Research}, Berkeley, 2007.}
\item{Biham, O., Middleton, A.A. \& Levine, D. 1992 Self-organization and a dynamical transition in traffic-flow models. \textit{Physical Review A}, \textbf{46} (10), 6124--6127.}
\item{Buisson, C. \& Ladier, C. 2009 Exploring the impact of the homogeneity of traffic measurements on the existence of macroscopic fundamental diagrams. \textit{Transportation Research Record} (in press).}
\item{Bretti, G., Natalini, R. \& Piccoli, B. 2007 Numerical algorithms for simulation of a traffic model on road networks. \textit{Journal of Computational and Applied Mathematics} \textbf{210}, 71--77.}
\item{Charypar, D. \& Nagel, K. 2005 Generating complete all-day activity plans with genetic algorithms. \textit{Transportation}, \textbf{32} (4), 369--397.}
\item{Daganzo, C.F. 1994 The cell transmission model: A dynamic representation of highway traffic consistent with the hydrodynamic theory. \textit{Transportation Research Part B: Methodological}, \textbf{28} (4), 269--287.}
\item{Daganzo, C.F. 2007 Urban gridlock: macroscopic modeling and mitigation approaches. \textit{Transportation Research Part B} \textbf{41} (1), 49--62. Corrigendum.
Transportation Research Part B 41 (3), 379.}
\item{Daganzo, C.F., \& Geroliminis, N. 2008 An analytical approximation for the macroscopic fundamental diagram of urban traffic. \textit{Transportation Research part B} \textbf{42} (9), 771--781.} 
\item{De Martino, D., Dall\'{}Asta, L., Bianconi, G. \& Marsili, M. 2009  A minimal model for congestion phenomena on complex networks. \textit{Journal of Statistical Mechanics: Theory and Experiment}, P08023.}
\item{Geroliminis, N. \& Daganzo, C.F. 2007 Macroscopic modeling of traffic in cities. In \textit{86th Annual Meeting of the Transportation Research Board}, Paper No. 07-0413, Washington, DC.}
\item{Geroliminis, N. \& Daganzo, C.F. 2008 Existence of urban-scale macroscopic fundamental diagrams: Some experimental findings. \textit{Transportation Research part B} \textbf{42} (9), 759--770.}
\item{Godfrey, J.W. 1969 The mechanism of a road network. \textit{Traffic Engineering and Control} \textbf{11} (7), 323--327.}
\item{Greenshields, B. D. 1935 A study in highway capacity. \textit{Highway Research Board, Proceedings} \textbf{Vol. 14}, 448--477.}
\item{Helbing, D. 2001 Traffic and related self-driven many-particle systems. \textit{Reviews of Modern Physics} \textbf{73}, 1067-1141.} 
\item{Helbing, D. 2003 A section-based queueing-theoretical traffic model for congestion and travel time analysis in networks. \textit{Journal of Physics A: Mathematical and General} \textbf{36} (46), 593--598.}
\item{Helbing, D. 2005 Production, supply, and traffic systems: A unified description. In \textit{Traffic and Granular Flow '03} (eds Hoogendoorn, S. P., Luding, S., Bovy, P. H. L. ,  Schreckenberg, M. \& Wolf, D. E.), 173--188, Springer, Berlin.}
\item{Helbing, D. 2009 Derivation of a fundamental diagram for urban traffic flow. \textit{The European Physical Journal B} \textbf{70} (2), 229--241.}
\item{Helbing, D. \& L\"{a}mmer, S. 2007 Method for coordination of concurrent competing processes for control of the transport of mobile units within a network. Pub. No. WO/2006/122528.}
\item{Helbing, D. \& Mazloumian, A. 2009 Operation regimes and slower-is-faster effect in the controlof traffic intersections. \textit{European Physical Journal B} \textbf{70} (2), 257--274.}
\item{Helbing, D. \& Moussaid, M. 2009 Analytical calculation of critical perturbation amplitudes and critical densities by non-linear stability analysis of a simple traffic flow model. \textit{European Physical Journal B} \textbf{69} (4), 571-581.} 
\item{Helbing, D., Siegmeier, J. \& L\"{a}mmer, S. 2007 Self-organized network flows. \textit{Networks and Heterogeneous Media} 
\textbf{2} (2), 193--210.}
\item{Helbing, D., Treiber, M., Kesting, A. \& Sch\"{o}nhof, M. 2009 Theoretical vs. empirical classification and prediction of congested traffic states. \textit{European Physical Journal B}   \textbf{69} (4), 583-598.}
\item{Herrmann, H.J. 1996 Spontaneous density fluctuations in granular flow and traffic. \textit{Nonlinear Physics of Complex Systems} (eds Parisi, J, M\"{u}ller, S.C. \& Zimmermann, W.), Lecture Notes in Physics No. 476, p. 23--34, Springer, Berlin.} 
\item{Herty, M., Pareschi, L. \& Seaid, M. 2006 Discrete-velocity models and relaxation schemes for traffic flows. \textit{SIAM Journal on Scientific Computing}, textbf{28}, 1582--1596.}
\item{Johansson, A., Helbing, D., Al-Abideen, H.Z. \& Al-Bosta, S. 2008 From crowd dynamics to crowd safety: A video-based analysis. \textit{Advances in Complex Systems}, \textbf{11} (4), 497--527.}
\item{Kerner, B. S. 1998 Experimental features of self-organization in traffic flow. \textit{Physical Review Letters}, \textbf{81} (17), 3797--3800.}
\item{Kerner, B. S. \& Rehborn, H. 1996 Experimental properties of complexity in traffic flow. \textit{Phys. Rev. E}, \textbf{53} (5), R4275--R4278.}
\item{Klar, A. \& Wegener, R. 1998 A hierarchy of models for multilane vehicular traffic I: modeling. \textit{SIAM Journal on Applied Mathematics}, 983--1001.}
\item{ L\"{a}mmer, S., Donner, R. \& Helbing, D. 2007 Anticipative control of switched queueing systems. \textit{The European Physical Journal B} \textbf{63} (3), 341--347.} 
\item{ L\"{a}mmer, S. \& Helbing, D. 2008 Self-control of traffic lights and vehicle flows in urban road networks. \textit{Journal of Statistical Mechanics: Theory and Experiment}, P04019.}
\item{Ma, T.Y. \& Lebacque, J.P. 2007 A cross entropy based multi-agent approach to traffic assignment problems. \textit{Proc. of the Traffic and Granular Flow}, 161--170, Springer, Berlin.}
\item{Mahmassani, H.S., Peeta, S. 1993 Network performance under system optimal and user equilibrium dynamic assignments: implications for ATIS. \textit{Transportation Research Record} \textbf{1408}, 83--93.}
\item{Mahmassani, H., Williams, J.C. \& Herman, R. 1987 Performance of urban traffic networks. In \textit{Proceedings of the 10th International Symposium on Transportation and Traffic Theory}  (eds Gartner, N.H. \& Wilson, N.H.M.), Elsevier, Amsterdam, The Netherlands.}
\item{Nagel, K. \& Schreckenberg, M. 1992 A cellular automaton model for freeway traffic. \textit{J. Phys. I France}, \textbf{Vol. 2}, 2221--2229.}
\item{Olszewski, P., Fan, H.S., Tan, Y.W. 1995 Area-wide traffic speed-flow model for the Singapore CBD. \textit{Transportation Research Part A} \textbf{29} (4), 273--281.}
\item{Orosz, G.,  Wilson, R. E., Szalai, R. \& St\'{e}p\'{a}n, G. 2009 Exciting traffic jams: Nonlinear phenomena behind traffic jam formation on highways. \textit{Physical Review E} \textbf{80} (4), 046205.}
\item{Padberg, K., Thiere, B., Preis, R. \&  Dellnitz, M. 2009  Local expansion concepts for detecting transport barriers in dynamical systems. \textit{Communications in Nonlinear Science and Numerical Simulations} \textbf{14} (12): 4176--4190.}
\item{Schadschneider, A. \& Schrenkenberg, M. 1993 Cellular automaton models and traffic flow. \textit{Journal of Physics-London-A Mathematical and General}, textbf{Vol. 26}, 679--683.}
\item{Tu, H. 2008 Monitoring travel time reliability of freeways. \textit{Transport \& Planning, Faculty of Civil Engineering and Geosciences}, 1--172.}
\item{Williams, J.C., Mahmassani, H.S. \& Herman, R. 1987 Urban traffic network flow models. \textit{Transportation Research Record} \textbf{1112}, 78--88.}
\item{Zhao, L., Lai, Y.C., Park, K. \& Ye, N. 2005 Onset of traffic congestion in complex networks. \textit{Physical Review E} \textbf{71} (2), 26125.}
\end{enumerate}
%-----------------------------------------------------------------------------------------------------------------------------------
\label{lastpage}
\end{document}